\documentclass[amsmath,amssymb,aps,prl,reprint]{revtex4-2}

\usepackage{aas_macros}
\usepackage{graphicx}
\usepackage{dcolumn}
\usepackage{bm}
\usepackage{hyperref}
\usepackage{soul}
\usepackage{xcolor}
\usepackage{cleveref}

\soulregister\cref7
\soulregister\ref7
\soulregister\cite7

\usepackage{CJK}
\crefname{figure}{Fig.}{Figs.}

\newcommand{\btheta}{{\bm \theta}}
\newcommand{\bx}{{\bf x}}

\begin{document}


\title{Cosmological Analysis with Calibrated Neural Quantile Estimation \\ and Approximate Simulators}

\author{He Jia (\begin{CJK*}{UTF8}{gbsn}贾赫\end{CJK*}) }
\email{hejia@princeton.edu}
\affiliation{Department of Astrophysical Sciences, Princeton University, Princeton, NJ 08544, USA}

\date{\today}

\begin{abstract}
A major challenge in extracting information from current and upcoming surveys of cosmological Large-Scale Structure (LSS) is the limited availability of computationally expensive high-fidelity simulations.
We introduce calibrated Neural Quantile Estimation (NQE), a new Simulation-Based Inference (SBI) method that leverages a large number of approximate simulations for training and a small number of high-fidelity simulations for calibration.
This approach guarantees an unbiased posterior regardless of approximate simulation accuracy, while achieving near-optimal constraining power when the approximate simulations are reasonably accurate.
As a proof of concept, we demonstrate that cosmological parameters can be inferred at field level from projected 2-dim dark matter density maps up to $k_{\rm max}\sim1.5\,h$/Mpc at $z=0$ by training on $\sim10^4$ Particle-Mesh (PM) simulations with transfer function correction and calibrating with $\sim10^2$ Particle-Particle (PP) simulations.
The calibrated posteriors closely match those obtained by directly training on $\sim10^4$ expensive PP simulations, but at a fraction of the computational cost.
Our method offers a practical and scalable framework for SBI of cosmological LSS, enabling precise inference across vast volumes and down to small scales.
\end{abstract}

\maketitle




\section{Introduction}

Current and upcoming spectroscopic and photometric surveys, such as DESI \citep{desi2022overview}, Euclid \citep{euclid2024overview}, Rubin \citep{lsst2019overview}, and Roman \citep{spergel2015roman}, will map the Large-Scale Structure (LSS) of our universe over vast volumes and down to unprecedented small scales, introducing significant challenges in cosmological parameter inference from the survey data.
On larger scales, the power spectrum serves as an optimal summary statistic and can be accurately calculated using perturbation theory.
However, on smaller scales, perturbation theory breaks down, and the power spectrum is no longer an optimal summary statistic, leaving us without a predefined statistic capable of fully capturing the complex information within cosmological density fields \citep{nguyen2024how,krause2024beyond}.

Simulation-Based Inference \citep[SBI,][]{cranmer2020frontier,lueckmann2021benchmarking}, also known as Likelihood-Free Inference (LFI) or Implicit Likelihood Inference (ILI), enables Bayesian inference directly from simulations, bypassing the need for an explicitly defined likelihood function.
Modern SBI methods include Neural Posterior Estimation \citep[NPE,][]{papamakarios2016fast,lueckmann2017flexible,greenberg2019automatic}, Neural Likelihood Estimation \citep[NLE,][]{papamakarios2019sequential,lueckmann2019likelihood}, Neural Ratio Estimation \citep[NRE,][]{hermans2020likelihood}, and the recently developed Neural Quantile Estimation \citep[NQE,][]{jia2024simulation}.
Typically, an SBI estimator is trained on simulated datasets $(\btheta, \bx)$, where $\btheta$ denotes model parameters and $\bx$ represents mock data --- either the full cosmological maps or their summary statistics for LSS applications.
For higher-dimensional $\bx$, an embedding network, such as a Convolutional Neural Network \citep[CNN,][]{lecun1989cnn,he2016resnet}, can help to extract relevant features from the data.
With sufficiently expressive neural networks and adequate training, SBI can infer the optimal posterior without requiring an explicit form of optimal summary statistics.

To accurately model the nonlinear regime of LSS, N-body simulations with Particle-Particle \citep[PP,][]{springel2021gadget,garrison2021abacus} interactions are required, and hydrodynamic simulations \citep{bryan2014enzo,weinberger2020arepo,stone2024athenak} become necessary at smaller scales where baryonic effects are significant; however, both approaches are computationally expensive.
Approximate methods, such as Particle-Mesh \citep[PM,][]{feng2016fastpm,li2022pmwd,rampf2024bullfrog} simulations and emulator-based models \citep{dai2018gradient,he2019learning,payot2023learning,sharma2024emulator,jamieson2024field}, offer faster alternatives and are typically validated by comparing their summary statistics to those from high-fidelity simulations.
These approximate simulators, however, cannot perfectly reproduce the results of exact simulations, introducing residual errors that propagate into inference results, making it challenging to
correct their impact on cosmological parameter estimates.
Consequently, state-of-the-art cosmological SBI analyses, such as SIMBIG \citep{lemos2024simbig}, continue to rely on stacking small-box PP simulations, as large-box PP or hydrodynamic simulations remain infeasible in the short term.
Unfortunately, super-sample effects are present in these small-box analyses, and while certain techniques attempt to account for them \citep{takada2013power,bayer2023super,modi2023hybrid}, these methods may still lack the accuracy and reliability needed for a robust and optimal cosmological analysis.

\begin{figure}[t]
    \includegraphics[width=0.45\textwidth]{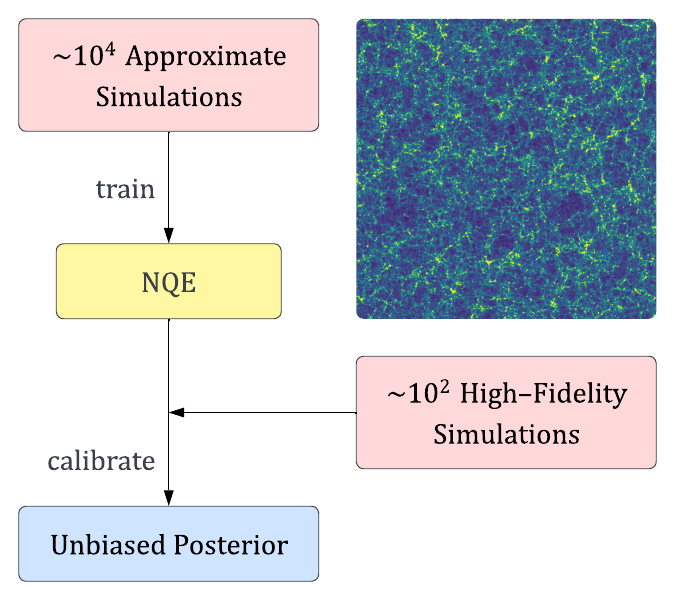}
    \caption{Flowchart of the proposed method, which first applies Neural Quantile Estimation (NQE) to infer cosmological parameters from cosmological maps using $\sim\!10^4$ approximate simulations, followed by calibration of the posterior with $\sim\!10^2$ high-fidelity simulations. An example 2-dim projected dark matter density field is shown in the upper right corner.}
    \label{fig:flow}
\end{figure}

In this work, we propose a framework to address this challenge by first training NQE on a large number of approximate simulations, followed by posterior calibration with a small number of high-fidelity simulations.
A flowchart of the proposed framework is shown in \cref{fig:flow}.
This approach guarantees an unbiased posterior, as evidenced by a diagonal empirical coverage curve, regardless of the accuracy of the approximate simulator.
The calibrated posterior is also close to optimal, compared with the posterior trained directly on high-fidelity simulations, as measured by the inverse volume of the credible regions, when the approximate simulator is reasonably accurate.
As a proof of concept, we demonstrate that cosmological parameters can be inferred from projected 2-dim dark matter density fields up to $k_{\rm max} \sim 1.5\,h$/Mpc at $z=0$ by training on $\sim10^4$ PM simulations (with transfer function correction \citep{sharma2024emulator}) and calibrating with $\sim10^2$ PP simulations.
This method yields posteriors (on mock data from PP simulations) comparable to those obtained by training directly on $\sim10^4$ PP simulations but with significantly lower computational cost.
Additionally, we find that a deep CNN, jointly trained with the inference network, consistently extracts more information than predefined summary statistics such as power spectra and scattering transform coefficients \citep{cheng2020scattering,valogiannis2022scattering,lin2024scattering}.
Our NQE algorithm is publicly available in the \texttt{nqe} package \footnote{\url{https://github.com/h3jia/nqe}}.

\section{Methods}

SBI methods enable direct inference of model parameters $\btheta$ from observational data $\bx$ without requiring an explicitly formulated likelihood function.
Given the simulated dataset $(\btheta, \bx)$, NQE \citep{jia2024simulation} first decomposes the joint posterior autoregressively into 1-dim conditional posteriors, $p(\btheta|\bx) = \prod_i p(\theta^{(i)} \,|\, \bx, \btheta^{(j<i)})$, and fits the quantiles of each $p(\theta^{(i)} \,|\, \bx, \btheta^{(j<i)})$ using neural networks by optimizing a weighted $L_1$ loss.
The full posterior $p(\btheta|\bx)$ is then reconstructed by interpolating the individual quantiles of $p(\theta^{(i)} \,|\, \bx, \btheta^{(j<i)})$.
Unlike NPE and NLE, NQE does not rely on Normalizing Flows \citep[NF,][]{rezende2015variational,papamakarios2021normalizing}, which allows for a simpler network architecture. It uses a fully connected Multi-Layer Perceptron (MLP) to predict quantiles, with an optional embedding network (e.g., a CNN) for high-dimensional observational data.

\begin{figure}[t]
    \includegraphics[width=0.48\textwidth]{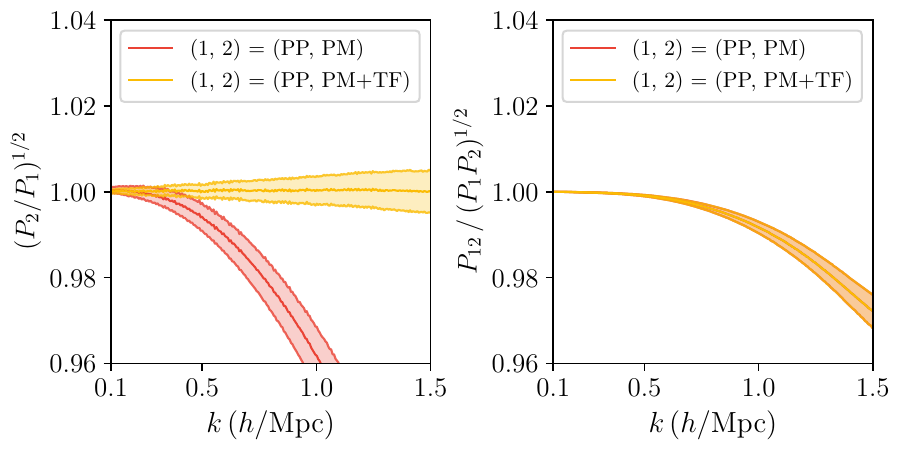}
    \caption{Comparison of power spectrum ratios and cross-correlation coefficients between the Particle-Particle (PP) simulation and two Particle-Mesh (PM) variants: the raw PM simulation and PM with transfer function correction (PM+TF). The PM+TF correction achieves sub-percent accuracy in the power spectrum up to $k \sim 1.5\,h$/Mpc but has no effect on the cross-correlation coefficient. Error bands represent the 25\%, 50\%, and 75\% quantiles across 100 realizations.}
    \label{fig:tf}
\end{figure}

One advantage of NQE over other SBI methods is its straightforward calibration.
In this work, we adopt a two-stage calibration strategy for NQE \citep{jia2024cali}.
The first stage, referred to as shift, adjusts each estimated $\tau$-th quantile to ensure it aligns with the $\tau$-th quantile on the calibration dataset, effectively averaging the quantile prediction errors over the calibration data.
The second stage, Importance Sampling (IS),
applies weights to each sample based on the rank of its probabilistic density, ensuring a diagonal empirical coverage curve.
While IS can be applied to other SBI methods, the shift step is exclusive to NQE.
Both steps are essential for effective calibration: the shift step alone does not guarantee correct posterior coverage \footnote{In this work, we adopt the standard definition of coverage based on the rank of posterior density, referred to as $p$-coverage in \citep{jia2024simulation}. However, the shift step alone does ensure correct $q$-coverage for each 1-dim conditional posterior.}, while the IS step alone may require an impractically large number of posterior samples and can result in a suboptimal posterior in terms of constraining power if the raw posterior is moderately biased.
A detailed description of NQE's training, calibration, and sampling procedures can be found in the Appendix.

\begin{figure*}[t]
    \includegraphics[width=\textwidth]{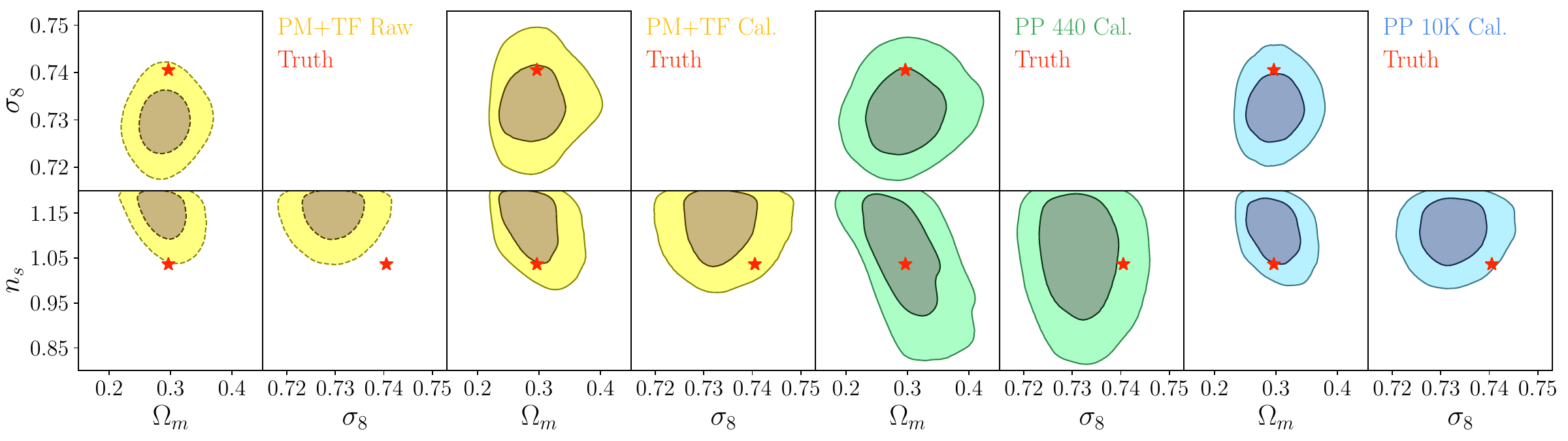}
    \caption{Posterior comparison for $k_{\rm max}\!=\!1.5\,h$/Mpc with CNN as the data compressor, trained on different simulations while tested on PP data. The uncalibrated estimator trained on PM+TF simulations is biased, but calibration effectively removes this bias. All calibrated estimators are unbiased; however, the calibrated PM+TF estimator achieves constraining power comparable to that of an estimator trained on $10^4$ PP simulations, and outperforms the estimator trained on 440 PP simulations.}
    \label{fig:sample}
\end{figure*}

We compare three approaches for extracting information from cosmological maps:
(1) PS, which uses the power spectrum of the image;
(2) ST+PS, which combines scattering transform coefficients \citep{cheng2020scattering,valogiannis2022scattering,lin2024scattering} up to the second order (with $J=8$ and $L=4$) and the power spectrum;
(3) CNN, where a deep residual network \citep[ResNet,][]{he2016resnet} is used to directly compress the information in the input maps.
Unlike prior works \citep{sharma2024comparative,lanzieri2024optimal,makinen2024hybrid}, the CNN is jointly trained with the inference network, as we find that separate training does not improve inference performance.
For PS and ST+PS, the statistics are directly fed into an MLP to estimate posterior quantiles, while for CNN, the raw images are first compressed by the CNN and then passed to the MLP.
We use the same MLP architecture and training procedure across all three methods to ensure a fair comparison of their information content.

\begin{figure*}[t]
    \includegraphics[width=\textwidth]{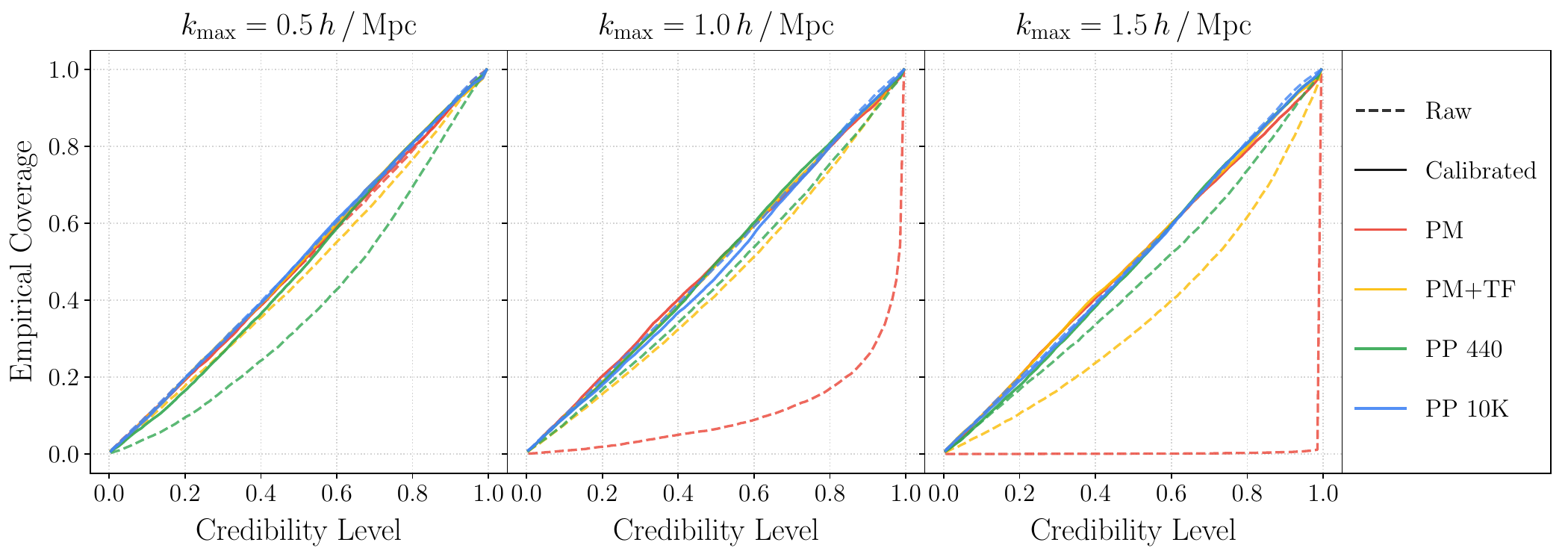}
    \caption{Empirical coverage of different estimators with CNN as data compressor. Uncalibrated estimators may exhibit bias due to limited training data (e.g., PP 440) or model misspecification (e.g., PM and PM+TF). By calibrating with 100 PP simulations, we achieve diagonal coverage curves across all methods, indicating unbiased posterior estimates.}
    \label{fig:cover}
\end{figure*}

As a proof of concept, we study the inference of $(\Omega_m, \sigma_8, n_s)$ from projected 2-dim dark matter overdensity maps.
For high-fidelity simulations, we use the PP simulations from the Quijote BSQ suite \citep{paco2020quijote}, where each simulation evolves $512^3$ particles in a 1 Gpc/$h$ box using the \texttt{GADGET} code \citep{springel2021gadget}.
For approximate simulations, we run \texttt{FastPM} \citep{feng2016fastpm} with the same parameter prior, box size, and particle number.
These PM simulations use a force resolution factor of $B=3$ and evolve over 25 time steps from $z=24$ to $z=0$, and are around 100 times faster than the PP simulations.
For each simulation, we compute the dark matter overdensity field at $z=0$ on a $512^3$ mesh, divide the box into eight slices along each axis, and project each slice into 2-dim, yielding 24 $512^2$ maps per simulation.
We apply a high-pass sharp-$k$ filter with $k_{\rm min}=0.1\,h/$Mpc and three low-pass filters with $k_{\rm max}=0.5\,h/$Mpc, $1.0\,h/$Mpc, and $1.5\,h/$Mpc to compare inference across different scales.
To correct for differences between PP and PM simulations, we use 50 pairs of simulations (with matching initial conditions) to train a Gaussian Process (GP) emulator for the transfer function, defined as the ratio of the power spectra between PP and PM maps, with the cosmological parameters as its inputs \citep{sharma2024emulator}.
This transfer function is then applied to the PM overdensity fields to generate corrected fields, referred to as PM+TF hereinafter.
\cref{fig:tf} presents a comparison of power spectra and cross-correlation coefficients between different simulations.

\section{Results}


\begin{figure*}[t]
    \includegraphics[width=\textwidth]{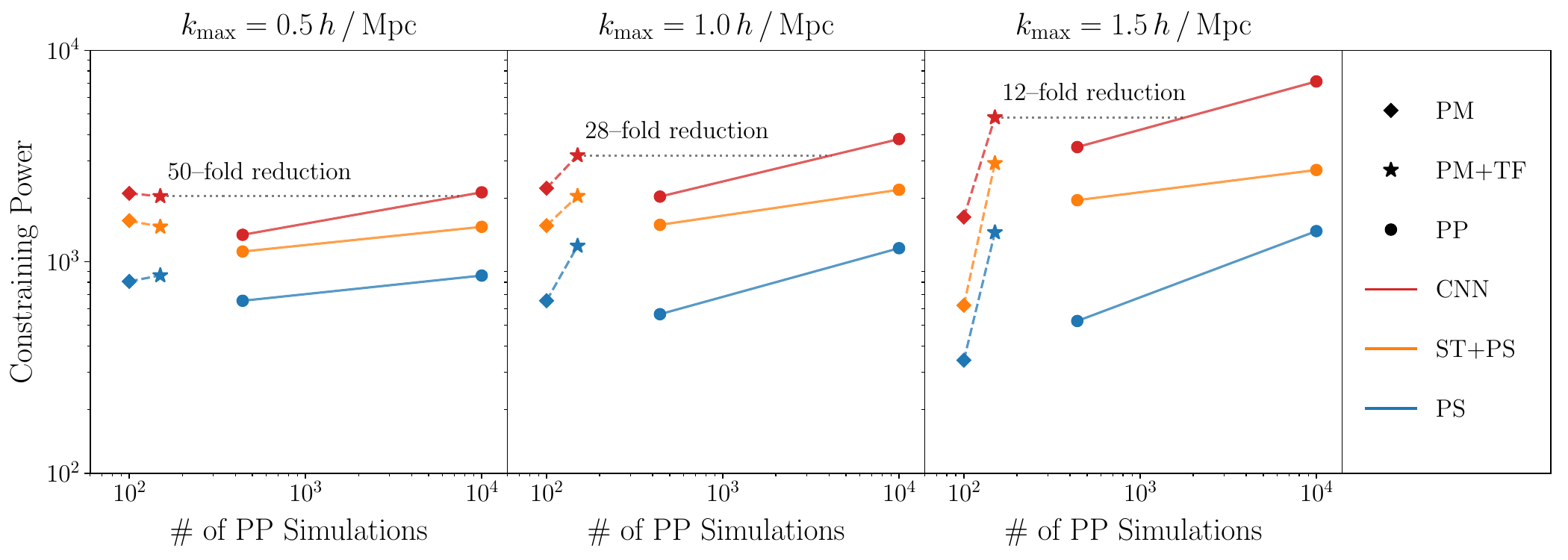}
    \caption{Comparison of the constraining power across different methods, as measured by the inverse volume of the 1-$\sigma$ credible region of the 3-dim posterior. All estimators have been calibrated to have a diagonal coverage curve (see \cref{fig:cover}). The proposed method, which utilizes PM+TF simulations to train the estimator and CNN to extract information, achieves comparable constraining power to estimators trained directly on a significantly larger set of PP simulations.}
    \label{fig:fom}
\end{figure*}

We train NQE with the three aforementioned data compression schemes (PS, ST+PS, and CNN) on four datasets: $10^4$ PM simulations without and with transfer function correction (referred to as PM and PM+TF, respectively), and 400 and $10^4$ PP simulations.
Each dataset is divided into 85\% for training and 15\% for validation, followed by a calibration step using 100 PP simulations.
For estimators trained on PP simulations, the validation dataset is reused for calibration, resulting in total PP simulation budgets of 440 and $10^4$ (referred to as PP 440 and PP 10K, respectively).
For PM+TF, we use 50 additional pairs of PM and PP simulations to train the GP transfer-function emulator with a radial basis function kernel, bringing the total PP simulation budget for this dataset to 150.
The PM and PM+TF datasets require additional $10^4$ PM simulations, with a computational cost comparable to $10^2$ PP simulations, which is not included in the PP simulation budgets.
We use an independent set of 500 PP simulations to evaluate the accuracy of the inferred posteriors.

In \cref{fig:sample}, we compare the posteriors obtained from estimators trained on different datasets.
The raw estimator trained on PM+TF data exhibits bias on PP data, but calibration effectively removes such bias.
The impact of calibration is further illustrated in \cref{fig:cover}, which presents the empirical coverage --- the probability that the true model parameters fall within the corresponding highest-posterior-density credible regions --- where a diagonal curve signifies a well-calibrated, unbiased estimator.
While the raw estimators may be biased due to insufficient training data (e.g., PP 440) or model misspecification (e.g., PM and PM+TF), calibration yields unbiased posteriors in all cases, as evidenced by the diagonal empirical coverage curves.

\cref{fig:sample} also shows that while all calibrated estimators are unbiased, their constraining power varies: the posterior contours from calibrated PM+TF are comparable to those from calibrated PP 10K but are noticeably tighter than those from PP 440.
A quantitative comparison of constraining power, as measured by the inverse volume of the 1-$\sigma$ credible region of the 3-dim posterior, is presented in \cref{fig:fom}.
All estimators have been calibrated to exhibit diagonal empirical coverage curves, ensuring a fair comparison, as the constraining power of overconfident estimators will be overestimated and vice versa.
We find that calibrated estimators trained on PM+TF achieve constraining power comparable to those trained on PP datasets that are 1--2 orders of magnitude more computationally expensive.
Additionally, CNN-based methods consistently demonstrate stronger performance than predefined summary statistics such as PS and ST, even in limited simulation-budget scenarios, although \citep{sharma2024comparative} found that for certain predefined summary statistics, Gaussian likelihood approximation may converge faster than SBI methods in such regimes.
These results underscore PM+TF combined with CNN as the most effective approach for extracting information from non-Gaussian cosmological maps under computational constraints.



\section{Discussion}

In this work, we present a framework to train SBI models using a large number of approximate simulations and calibrate them with a small number of high-fidelity simulations.
We demonstrate its effectiveness by addressing the problem of inferring cosmological parameters from projected 2-dim dark matter density fields.
To the best of our knowledge, this is the first work to show that approximate simulators, such as PM+TF, can be effectively calibrated with a limited number of high-fidelity simulations, enabling unbiased inference down to small scales where the approximate simulators exhibit reduced accuracy.
The proposed framework is broadly applicable to any problem involving expensive high-fidelity simulators paired with faster approximate counterparts.
Without such calibration, the utility of approximate simulators for generating SBI training data is binary: they are either accurate enough or not, based on the error tolerance.
Calibration transforms this binary evaluation into a continuum, ensuring that the posterior remains unbiased and therefore always ``usable", while the accuracy of the approximate simulator primarily impacts the constraining power of the posterior.

Although we apply this method specifically to the inference of 3-dim parameters $(\Omega_m, \sigma_8, n_s)$, we do not expect the number of high-fidelity simulations required for calibration to depend strongly on the dimensionality of $\btheta$.
In our two-stage calibration strategy, the initial shift step is applied independently to each 1-dim conditional posterior, after which the empirical coverage curve is typically close to diagonal (see \cref{fig:cover-is-shift}).
As a result, the IS weights will be near unity, requiring relatively few simulations for the IS step to converge.
In this example, each simulation generates 24 maps, which are not fully independent; consequently, the effective number of independent maps for calibration likely falls between 100 and 2400.
The number of simulations needed for calibration therefore also depends on the number of independent observational data points that can be produced per simulation.
Nonetheless, we expect that approximately $\mathcal{O}(10^2)$ simulations should be sufficient for calibration in most scenarios, except when $\btheta$ is high-dimensional with complex correlations between dimensions, leaving empirical coverage curves significantly off-diagonal even after the initial shift step.


The information loss due to the calibration process, reflected in the difference in constraining power between calibrated estimators trained on approximate simulations and those trained on the same number of high-fidelity simulations, depends on the accuracy of the approximate simulator.
In this work, we apply a simple transfer function corrector \citep{sharma2024emulator} to the PM simulations, achieving good accuracy for the dark matter density fields up to $k\sim1.5\,h$/Mpc at $z=0$.
This level of accuracy is likely sufficient for most current and upcoming spectroscopic and weak lensing surveys, except for applications involving the very local universe, where small-scale structures beyond $k\sim1.5\,h$/Mpc may also influence the observables.
For simplicity, we focus on real-space dark matter density fields, which can be directly corrected in Eulerian space using the transfer function emulator.
However, for practical applications involving halo models and/or redshift-space observables, Lagrangian-space emulators \citep{dai2018gradient,he2019learning,payot2023learning,jamieson2024field} may be more appropriate, as they can provide reasonably accurate predictions of particle positions and velocities.
With such advanced approximate simulators, we anticipate the ability to perform near-optimal cosmological inference down to even smaller scales.



While this work focuses on inference performance under the assumption that the high-fidelity simulations perfectly represent real-world data, we do not address the robustness of the estimators in the presence of ``unknown" systematics, a question we leave for future research. 
We do, however, demonstrate that while minor ``known" inaccuracies in the simulations (e.g., the PM+TF dataset) may affect the posterior estimation based on CNN or predefined summary statistics, these effects can be effectively corrected through calibration.



We note several recent efforts aimed to adapt inference algorithms trained on one simulator for application to another, either by selecting subsets of observables that remain relatively consistent across simulators \citep{santi2023robust,rojas2023cosmology} or by employing domain adaptation techniques during training \citep{roncoli2023domain,lee2024inferring}.
However, these approaches cannot guarantee unbiasedness, as indicated by the absence of a diagonal empirical coverage curve under limited high-fidelity simulation budgets, nor is it clear whether substantial information loss occurs during the adaptation process.
Hybrid SBI \citep{modi2023hybrid}, in contrast, models large-scale observables with perturbation theory and small-scale observables with simulations, thus avoiding the need for large-box high-fidelity simulations.
Nevertheless, this method requires modeling the likelihood of small-scale observables conditioned on large-scale observables, a task that remains challenging without actually running large-box simulations.
Furthermore, on small scales, their approach relies on predefined summary statistics such as scattering transform coefficients, which, as shown in this work, are suboptimal compared to CNNs.

Building on this work, we plan to run large-box approximate simulations of cosmological LSS for SBI analyses of current and upcoming surveys, enabling us to extract deeper insights from the data that would otherwise remain inaccessible due to practical computational limitations.

\begin{acknowledgments}
We thank Adrian Bayer, Sihao Cheng, Biwei Dai, ChangHoon Hahn, Nickolas Kokron, Peter Melchior, Uro\v{s} Seljak, David Spergel, and Hongming Zhu for helpful discussions.
The work presented in this article was performed on computational resources managed and supported by Princeton Research Computing, a consortium of groups including the Princeton Institute for Computational Science and Engineering (PICSciE) and the Office of Information Technology's High Performance Computing Center and Visualization Laboratory at Princeton University.
\end{acknowledgments}


\bibliography{apssamp}

\appendix

\section{The Neural Quantile Estimation Algorithm}

NQE \citep{jia2024simulation} infers model parameters $\btheta$ from observational data $\bx$ by estimating approximately 20 individual quantiles for each 1-dim conditional distribution $p(\theta^{(i)} \,|\, \bx, \btheta^{(j<i)})$.
These quantiles are predicted using neural networks optimized with a weighted $L_1$ loss.
The full posterior distribution is then reconstructed by interpolating the predicted quantiles.
We employ a hybrid interpolation scheme: the Cumulative Distribution Function (CDF) of each 1-dim conditional distribution is interpolated using cubic splines, except in regions identified as tails, where Gaussian tails are used to mitigate interpolation artifacts.
Posterior samples are generated by sequentially sampling from the 1-dim conditional distributions, $\theta^{(i)} \sim p(\theta^{(i)} \,|\, \bx, \btheta^{(j<i)})$.
We use a separate neural network to fit the quantiles of each posterior dimension, while the training for different dimensions is independent, enabling parallelization across multiple GPUs.

We adopt a two-stage calibration strategy for NQE \citep{jia2024cali}.
In the first stage, referred to as shift calibration, we assess the residual between the true $\theta^{(i)}$ and the predicted $\tau$-th quantiles of the 1-dim conditional posterior, evaluated at the true model parameters.
Ideally, the $\tau$-th quantile of these residuals across the calibration dataset should be zero, indicating that the predicted $\tau$-th quantile aligns with the actual $\tau$-th quantile.
If this condition is not met, we apply a shift correction to the predicted quantiles to compensate for the error.
This correction depends on $\tau$ and the posterior dimension index $i$, but is independent of the specific $\btheta$ and $\bx$, effectively averaging the quantile prediction errors across the entire calibration dataset. 
To address hard prior boundaries of $\btheta$, we first apply a logit transform to map the parameters to an unconstrained space, perform the shift calibration, and then transform the parameters back.

After the shift calibration, the posterior is typically close to unbiased, but there is no guarantee of perfectly diagonal empirical coverage.
To address this, we apply a second IS step.
For each calibration data point, we compute the quantile of the posterior probability density at the true $\btheta$ among 2048 posterior samples, and then construct a histogram of these quantiles using 16 bins.
A perfectly calibrated estimator would yield a flat histogram, indicating uniform empirical coverage.
If deviations from flatness are observed, IS weights are applied to the posterior samples, determined by the rank of their probabilistic density, to correct the deviation and ensure that the calibrated posterior achieves accurate empirical coverage.

\section{Network Architecture and Training Details}

\begin{figure}[h]
    \includegraphics[width=0.4\textwidth]{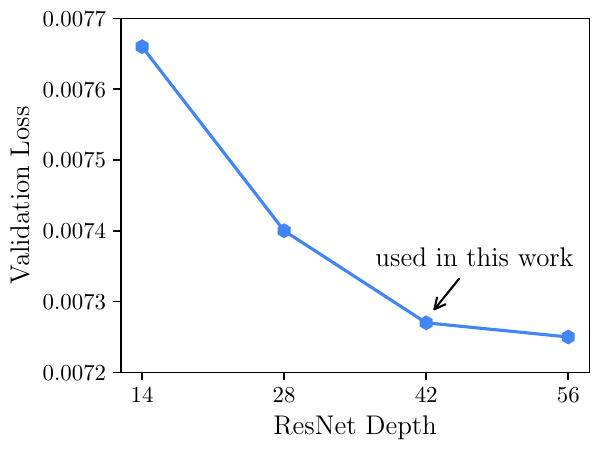}
    \caption{Validation loss for $\Omega_m$, evaluated on PP simulations with $k_{\rm max} = 1.5\,h$/Mpc, using a CNN as data compressor, as a function of CNN depth. The loss decreases with increasing depth and converges at $\gtrsim42$ layers. Accordingly, we use a 42-layer CNN in this work to compress information from cosmological maps.}
    \label{fig:loss}
\end{figure}

We use a modified version of the original ResNet architecture \citep{he2016resnet} for direct, field-level data compression.
Given a $512^2$ image, the network consists of seven sections of convolutional layers, each comprising $N_{\rm block}$ Basic Blocks followed by a mean pooling layer that downsamples the image resolution by a factor of 2.
The number of channels in each section is set to 50, 100, 100, 200, 400, 400, and 800, respectively.
After the final convolutional layer, we apply an additional mean pooling layer to reduce the $800\times4\times4$ output to a flat array of 800 features.
Each Basic Block contains two convolutional layers, resulting in a total of $14\,N_{\rm block}$ layers for the ResNet.
To evaluate the network depth required for effective inference, we test the convergence of inference performance with respect to the number of layers.
As shown in \cref{fig:loss}, our 42-layer ResNet adequately captures the majority of the information within the cosmological maps.

Given the 800 features extracted by the ResNet or the predefined summary statistics directly calculated from the images, we use a 10-layer MLP with 800 hidden neurons per layer and concatenation shortcuts to predict the quantiles.
We find the inference results to be relatively insensitive to reasonable variations in the MLP depth and width.
The network is trained using the Adam optimizer \citep{ba2015adam} with an initial learning rate of $10^{-4}$ for 10 epochs, followed by 50 additional epochs with the learning rate reduced by 10\% at each epoch. 
For the PP 440 dataset, we instead train for 40 initial epochs, followed by 200 epochs with the learning rate decayed by 2.5\% per epoch.
We adopt the default settings recommended by \citep{jia2024simulation} for other hyperparameters of NQE and use early termination to mitigate potential overfitting.
While predefined summary statistics, such as scattering transform coefficients, are less optimal than CNNs in terms of constraining power, we find that training with them is significantly faster.
For example, training the CNN-based networks on the $10^4$-simulation dataset requires approximately 20 hours on 4 Nvidia A100 GPUs per posterior dimension, whereas the same training with predefined summary statistics takes less than 30 minutes on a single Nvidia MIG instance, which has computational power roughly equivalent to $1/7$ of an A100 GPU.

We evaluate the empirical coverage and constraining power of the various estimators using a test dataset of 500 independent PP simulations.
For the constraining power, we uniformly draw 4096 samples from a cube defined by $\left[\,{\bm q}_{0.5}-1.5\times({\bm q}_{0.5}-{\bm q}_{0.05}),\,{\bm q}_{0.5}+1.5\times({\bm q}_{0.95}-{\bm q}_{0.5})\,\right]$, where the $\bm q$ values represent the quantiles of the 1-dim marginal posteriors.
We then calculate the fraction of these samples that fall within the 1-$\sigma$ credible regions, as defined by the rank of posterior density.

\begin{figure*}[t]
    \includegraphics[width=\textwidth]{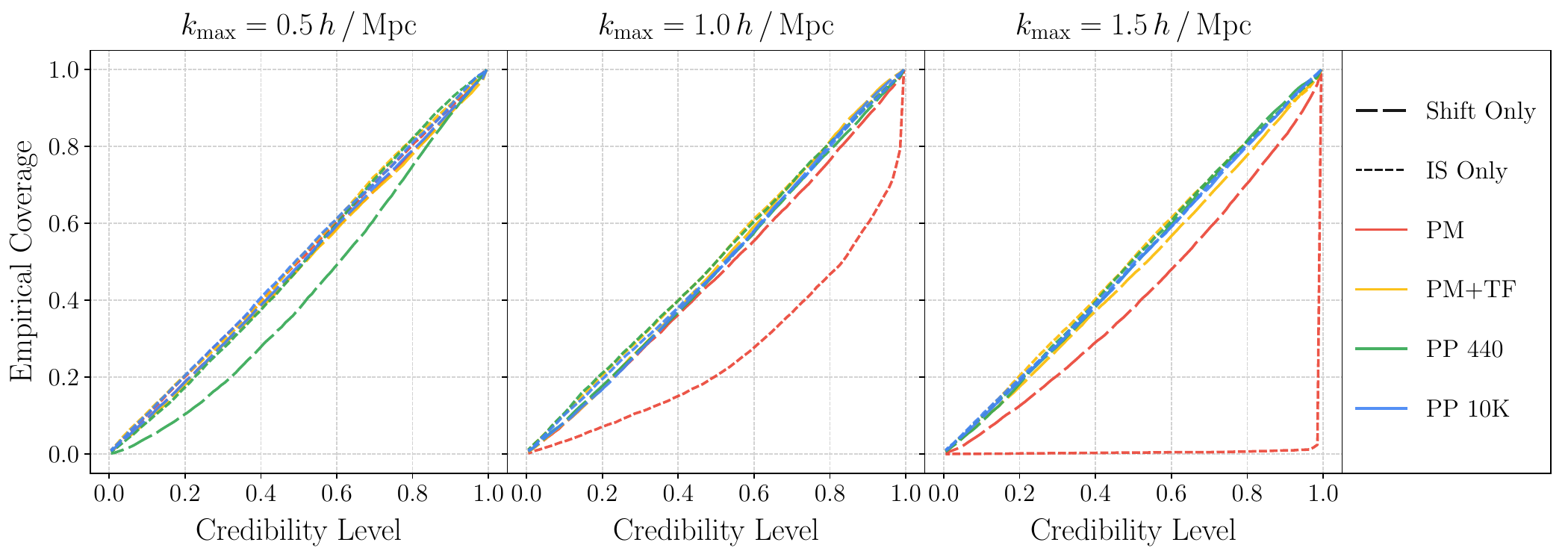}
    \caption{Similar to \cref{fig:cover}, but comparing estimators calibrated using only quantile shifting (Shift Only) and only importance sampling (IS Only). Both calibration steps are necessary to achieve accurate coverage across all cases.}
    \label{fig:cover-is-shift}
\end{figure*}

\begin{figure}[h]
    \centering
    \begin{minipage}{0.24\textwidth}
        \centering
        \includegraphics[width=\linewidth]{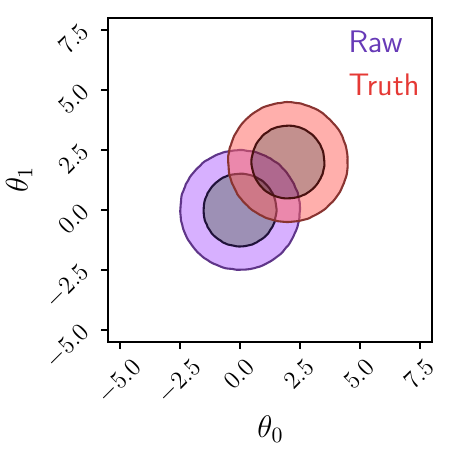}
    \end{minipage}%
    \hfill
    \begin{minipage}{0.24\textwidth}
        \centering
        \includegraphics[width=\linewidth]{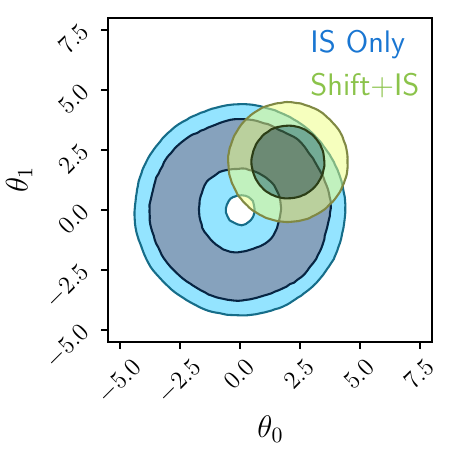}
    \end{minipage}
    \caption{A toy example illustrating the limitations of IS-only calibration. When the uncalibrated posterior is systematically biased, IS-only calibration becomes suboptimal, distorting the Gaussian posterior into a ring shape to fix the coverage curve, as it cannot modify the isodensity surfaces of the posterior. In contrast, the two-stage calibration effectively recovers the Bayesian-optimal posterior.}
    \label{fig:gaussian}
\end{figure}

\section{Additional Discussion on Calibration}


In \cref{fig:cover-is-shift}, we compare calibration strategies that use only the shift step or only the IS step, neither of which achieves diagonal coverage curves across all cases.
When using only the shift step, there is no guarantee of diagonal empirical coverage, although the deviations are generally small.
With IS alone, calibration fails when the raw posterior is significantly biased.
This breakdown is largely due to an insufficient number of posterior samples generated during calibration:
in such cases, the posterior evaluated at the true model parameters is likely smaller than the majority, if not all, of the 2048 samples drawn from the uncalibrated posterior, necessitating a significantly larger sample set to accurately capture the rank dependence of the IS weights.
While this does not necessarily increase the number of simulations needed for calibration, it may pose a challenge if an excessively large number of posterior samples must be drawn for the process.

\begin{figure}[h]
    \includegraphics[width=0.4\textwidth]{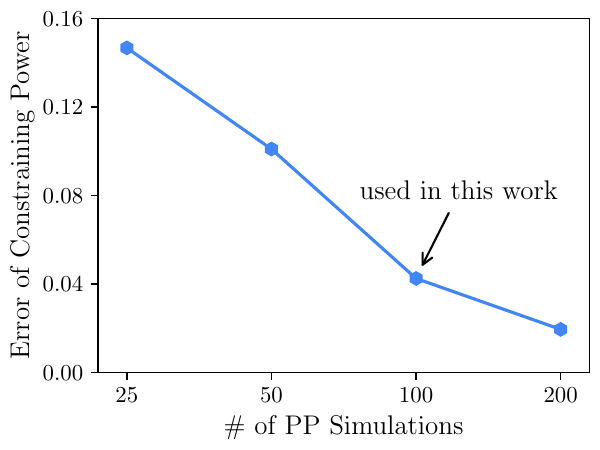}
    \caption{Convergence of constraining power (as defined in \cref{fig:fom}) as a function of the number of PP simulations used for calibration. For each scenario shown in \cref{fig:fom}, we calculate the relative error in constraining power with respect to a baseline obtained from 500 PP simulations, and then average over all 36 scenarios. With $\gtrsim 10^2$ PP simulations, the relative error falls below $\sim 5\%$.}
    \label{fig:fomcov}
\end{figure}

The effects of IS-only calibration are further illustrated with the toy example in \cref{fig:gaussian}: we assume the true posterior is a 2-dim Gaussian centered at (2, 2) with unit variance, while the raw uncalibrated posterior is systematically biased by a shift of (--2, --2).
Such bias can be perfectly corrected using the two-stage Shift+IS calibration.
However, with IS alone, unbiasedness is achieved by distorting the Gaussian posterior into a ring, as IS cannot modify the isodensity surfaces of the posterior.
This IS-only approach is not only suboptimal in terms of constraining power but also computationally expensive, as it requires a large number of posterior samples to account for the significantly upweighted far tails of the uncalibrated posterior.
For example, generating \cref{fig:gaussian} requires $10^8$ samples, posing a challenge for realistic applications where posterior sampling is slower than in this toy example.

In this work, the PM+TF setup uses a total of 150 PP simulations: 50 to train the transfer-function emulator and 100 for posterior calibration. For some applications, however, even $\sim 10^2$ high-fidelity simulations may be computationally challenging, motivating an assessment of whether fewer calibration simulations could suffice.
Because the transfer-function emulator is specific to this application, we keep the 50 simulations used for emulator training fixed, and study the convergence of the calibration step with respect to the number of PP calibration simulations.
\cref{fig:fomcov} shows the resulting error in constraining power, evaluated with respect to a baseline using 500 PP simulations and averaged over all 36 scenarios.
We find that with $\gtrsim 10^2$ PP simulations, the relative error in constraining power falls below $\sim 5\%$, supporting our default choice of 100 PP simulations for calibration. Using fewer PP simulations is also possible, although the increased calibration error signifies a possible residual bias in the calibrated posterior.



\end{document}